\documentclass[twocolumn,showpacs,preprintnumbers,amsmath,amssymb,APSl,prd,nofootinbib,superscriptaddress]{revtex4-1}

\usepackage{bm}
\usepackage{mathrsfs}
\usepackage{xcolor,color,graphicx,graphics}
\usepackage[all]{xy}
\usepackage{epsfig,subfigure}
\usepackage{latexsym,amssymb,amsmath,amsfonts} 
\usepackage[english]{babel} 
\usepackage[OT1]{fontenc}
\usepackage[latin1]{inputenc}
\usepackage{makeidx}
\usepackage{hyperref}
\usepackage{color,graphicx,graphics,wrapfig,epsf}

\definecolor{red}{rgb}{1,0,0}

\def\+{^\dagger}

\def\<{\leftarrow}
\def\>{\rightarrow}

\def\({\left(}
\def\){\right)}

\newcommand{\bi}{\begin{itemize}} 				\newcommand{\ei}{\end{itemize}}
\newcommand{\benu}{\begin{enumerate}} 		\newcommand{\enu}{\end{enumerate}}
\newcommand{\bd}{\begin{dinglist}{0}}     \newcommand{\ed}{\end{dinglist}}
\newcommand{\bfig}{\begin{figure}[htbp]}  \newcommand{\efig}{\end{figure}}
        			
\newcommand{\bc}{\begin{center}} 				  \newcommand{\ec}{\end{center}}
\newcommand{\be}{\begin{equation}} 				\newcommand{\ee}{\end{equation}}
\newcommand{\bsub}{\begin{subequations}}  \newcommand{\esub}{\end{subequations}}
\newcommand{\ben}{\begin{eqnarray}} 			\newcommand{\een}{\end{eqnarray}}
\newcommand{\ba}[1]{\begin{array}{#1}} 		\newcommand{\ea}{\end{array}}
\newcommand{\bea}{\begin{equation}\begin{array}{rcl}}
\newcommand{\eea}{\end{array}\end{equation}}

\begin{document}
\title{Early evolutionary tracks of low-mass stellar objects in modified gravity}

\author{Aneta Wojnar}
\email{aneta.magdalena.wojnar@ut.ee}
\affiliation{Laboratory of Theoretical Physics, Institute of Physics, University of Tartu,
W. Ostwaldi 1, 50411 Tartu, Estonia
}

\begin{abstract}
Using a simple model of low-mass stellar objects we have shown modified gravity impact on their early evolution, such as Hayashi tracks, radiative 
core development, effective temperature, masses, and luminosities. We have also suggested that the upper mass' limit of fully convective stars
on the Main Sequence might be different than commonly adopted.
\end{abstract}

\maketitle

\section{Introduction}
Working on modified gravity does not make one to forget the elegance and success of Einstein's theory of gravity, being already confirmed by many
observations \cite{Will:2014kxa}; even more, General Relativity (GR) still delights when one of its mysterious predictions, such as the existence 
of black holes, is directly affirmed by the finding of gravitational waves as a result of black holes' binary
mergers \cite{TheLIGOScientific:2017qsa} as well as soon after the imaging of the shadow 
of the supermassive black hole of M87 \cite{Akiyama:2019cqa}
(see \cite{Barack:2018yly} for a review).

Despite the successes, there are still shortcomings which, among many others, the modified gravity community is trying to solve by proposing extensions or/and alternatives  
to GR. The lack of detection of dark 
matter and dark energy sources in order to be in agreement with the cosmological concordance
model \cite{Copeland:2006wr,Nojiri:2006ri,nojiri2,nojiri3,Capozziello:2007ec,Carroll:2004de}, issues with unification with the high energy physics
 \cite{ParTom,BirDav} as well as existence of space-time singularities \cite{Senovilla:2014gza} are just sequent motivations for modified theories of 
gravity.

 In order to answer the mentioned ambiguities, some of the extensions of GR propose to consider more general than Einstein-Hilbert gravitational action 
 (with non-linear terms of curvature scalars) \cite{DeFelice:2010aj}, or to include minimally or non-minimally coupled
scalar fields \cite{brans, Bergmann}, to incorporate additional geometric ingredients \cite{BeltranJimenez:2019tjy}, and to treat the physical 
constants as dynamical quantities  \cite{Dabrowski:2012eb, Leszczynska:2014xba}, as a few examples.
Some of those theories can be already constrained by gravitational wave 
observations \cite{Ezquiaga:2017ekz,Baker:2017hug,Creminelli:2017sry,Langlois:2017dyl,Sakstein:2017xjx,Lombriser:2015sxa}.

Astrophysics also provides additional constraints (one of the most exciting is the neutron stars' merger \cite{TheLIGOScientific:2017first})
on GR and its modifications \cite{Berti:2015itd,sun1,sun2}, but also delivers 
drawbacks. The
observations of neutron stars (NS) with masses of two solar ones \cite{lina,as,craw} clearly demonstrate how much is not understood yet when we try to 
construct a NS model with our current knowledge on gravitational interactions and nuclear physics at the extreme densities. The lack of information on 
the conditions at the NS center, where densities are above the nuclear saturation density $\rho\approx2.8\times 10^{14}\text{gr}/\text{cm}^3$, forces
to 
extrapolate forms of equation of state (EoS). The recent discovery of an object of $2.6$ solar masses \cite{NSBH} (being above the heaviest known 
NS and below the lightest black hole, that is, in the so-called mass gap \cite{straight}), which merged with a black hole of $23M_\odot$, provided even more
questions for theoretical physics of compact objects.

However, it turns out that there is a class of stellar objects, with the internal structure much better understood than that of neutron stars, 
which might be used to constrain theories of gravity. It is a family of low-mass stars (LMS) \cite{chab,chab3, lau} which includes such ordinary
objects as M dwarfs (also called red dwarfs), which are cool Main Sequence stars with masses in the range $[0.09-0.6]M_\odot$, brown dwarfs 
(stars which failed to join the Main Sequence, with masses below $0.09M_\odot$) \cite{kumar, Burrows:1992fg} or infant stars, 
that is, pre-main-sequence (PMS) stars which have not yet started hydrogen burning \cite{bert}. 

LMS are the most common stellar objects - around 70 percent of stars in the Milky Way are red dwarfs, evolving slowly because of their small size and 
low masses. Their importance becomes clear \cite{catelan} when one wants to understand the properties of distant galaxies - a significant 
part of baryonic mass is contained in LMS \cite{kroupa}; globular clusters, being the oldest objects in the Universe with reliable age determination,
mainly consists of such objects \cite{sand}. Another, maybe the most important, argument for studying LMS is our relationship with the Sun: the
knowledge on its past and future, that is, the evolution of a star with dependent planets like the Earth, crucially determines our 
fate and survival possibilities. There are already many discoveries \cite{cass,ali,harps1,harps2,yun,brad,ramses,bohn} of exoplanets orbiting low-mass stars whose habitable properties
essentially depend on star's characteristics, from which the most important one is its mass. Moreover,
these objects are also studied in order to test properties of Standard Model particles as well as dark matter
candidates \cite{raf1,raf2,vieira}.There are also works constraining the gravitational theories using data obtained from the Sun and white dwarfs observations \cite{helio,kum}, such as, for instance, analyzing the effect of an
additional Yukawa component for gravity in the stellar equilibrium \cite{orf1},
the role of a non-minimal coupling between matter and curvature \cite{orf2} or the effect of an ungravity component \cite{orf3, orf4}.

The early evolution of low-mass stars has not been yet examined in the context of modified gravity according to our knowledge (for a recent review on 
stellar structure in modified gravity see \cite{reva} while on astrophysical test see \cite{bak}. Main Sequence and red giant stars in modified gravity were studied in \cite{chang,davis,chow}.). Due to that fact, we would like to answer the questions if modified gravity may have any significant 
effect on early life of such objects as well as what kind of consequences might be carried by the positive answer to that issue. Accordingly, we will focus on the 
simplest example of metric-affine theories of gravity, that is, Palatini $f(\mathcal{R})$ gravity; nevertheless, the analogous analysis can be made for any 
other theory of gravity which modifies the Newtonian hydrostatic equilibrium and related to that stellar equations.

The action of Palatini $f(\mathcal{R})$ gravity, which is the simplest generalization of GR, has the following form
\begin{equation}
S=S_{\text{g}}+S_{\text{m}}=\frac{1}{2\kappa}\int \sqrt{-g}f(\mathcal{R}) d^4 x+S_{\text{m}}[g_{\mu\nu},\psi_m],\label{action}
\end{equation}
where $\mathcal{R}=\mathcal{R}^{\mu\nu}g_{\mu\nu}$ is the Ricci scalar constructed with the metric $g$ and Ricci tensor built of the independent 
connection $\hat\Gamma$. Thus the common assumption on $g$-metricity of the connection is abandoned. Let us notice that we use the $(-+++)$ metric signature convention while $\kappa=-\frac{8\pi G}{c^4}$ \cite{weinberg}.
The variation of \ref{action} with respect to the metric $g_{\mu\nu}$ gives
\begin{equation}
f'(\mathcal{R})\mathcal{R}_{\mu\nu}-\frac{1}{2}f(\mathcal{R})g_{\mu\nu}=\kappa T_{\mu\nu},\label{structural}
\end{equation}
where $T_{\mu\nu}$ is the energy 
momentum tensor of the matter field, obtained in the standard way $T_{\mu\nu}=-\frac{2}{\sqrt{-g}}\frac{\delta S_m}{\delta g_{\mu\nu}}$.
Later on it will be assumed to be a perfect fluid. Here, the primes denote derivatives with respect to the 
function's argument: $f'(\mathcal{R})=\frac{df(\mathcal{R})}{d\mathcal{R}}$.

The variation with respect to the independent connection $\hat\Gamma$ provides
\begin{equation}
\hat{\nabla}_\beta(\sqrt{-g}f'(\mathcal{R})g^{\mu\nu})=0.\label{con}
\end{equation}
We immediately notice that $\hat{\nabla}_\beta$ is the covariant derivative calculated with respect to $\hat\Gamma$, that is, it is the Levi-Civita connection 
of the conformal metric
\begin{equation}\label{met}
h_{\mu\nu}=f'(\mathcal{R})g_{\mu\nu}.
\end{equation}
A very helpful equation, called the structural equation, is obtained from the trace of \ref{structural} taken with respect
to $g_{\mu\nu}$
\begin{equation}
f'(\mathcal{R})\mathcal{R}-2 f(\mathcal{R})=\kappa T,\label{struc}
\end{equation}
where $T$ is the trace of the energy-momentum tensor. For some chosen functional $f(\mathcal R)$
it is possible to solve the structural equation \ref{struc} in order to obtain the relation $\mathcal{R}=\mathcal{R}(T)$. 

It is a well-known fact, derived easily from \ref{struc}, that in the vacuum the Palatini gravity provides Einstein vacuum solution with the cosmological constant independently of the $f(\mathcal{R})$ form. Moreover, in the case of analytic $f(\mathcal{R})$ functions it was shown \cite{junior} that the center-of-mass orbits are the same as in GR while the modifications of energy and momentum which appear in Euler equation are not sensitive to the experiments performed for the solar system orbits so far. The situation may change when atomic level experiments will be 
available \cite{sch,ol1,ol2}.

 It can be shown \cite{DeFelice:2010aj} that one may rewrite the field equations as a dynamical equation for the conformal 
 metric $ h_{\mu\nu}$ \cite{BSS,SSB} and the undynamic scalar 
 field
 defined as $\Phi=f'(\mathcal{R})$:
 \begin{subequations}
	\begin{align}
	\label{EOM_P1}
	 \bar R_{\mu\nu} - \frac{1}{2} h_{\mu\nu} \bar R  &  =\kappa \bar T_{\mu\nu}-{1\over 2} h_{\mu\nu} \bar U(\Phi)
	\end{align}
	\begin{align}
	\label{EOM_scalar_field_P1}
	  \Phi\bar R &  -  (\Phi^2\,\bar U(\Phi))^\prime =0
	\end{align}
\end{subequations}
where we have introduced $\bar U(\Phi)=\frac{\mathcal{R}\Phi-f(\mathcal{R})}{\Phi^2}$ and appropriate energy momentum 
tensor $\bar T_{\mu\nu}=\Phi^{-1}T_{\mu\nu}$. It has been already shown \cite{aneta,o,o1,o2} that this representation of 
the Palatini gravity significantly simplifies considerations on particular physical problems.

\section{Palatini stars}
The stellar structure in the metric-affine theory (for the detailed review on that topic see \cite{reva}) was studied mainly in the context
of spherical-symmetric solutions and mass-radius
relation \cite{mr1,mr2,mr3, mr4, mr5,mr6,mr7,mr8,mr9,mr10},
the last one given by the modified Tolman-Oppenheimer-Volkoff equations.
Possible issues and their solutions were discussed in \cite{ba1,ba2,ba3,pani,ba4,ba5,gonzalo2,kim,b6,gd}. Works on
stability problems can be found in \cite{aneta,stab1,stab2,stab3,stab4,stab5}. Non-relativistic stars which are our concern were considered 
in \cite{aneta2,gonzalo,artur,nn1,nn2,aneta3}. In what follows, we will use some results derived in \cite{aneta,aneta2,gonzalo,artur}.

\subsection{Non-relativistic Palatini stars}\label{NRs}

It was demonstrated \cite{aneta2,gonzalo} for the Starobinski model 
\begin{equation}
 f(\mathcal{R})=\mathcal{R}+\beta\mathcal{R}^2
\end{equation}
that non-relativistic Palatini stars can be described by the following equations 
\begin{align}
 \frac{dp}{d\tilde r}&=-\frac{G m(\tilde r)\rho(\tilde r)}{\Phi(\tilde r)\tilde r^2} \ ,\\
 m&=\int_0^{\tilde r}4\pi x^2\rho(x)dx \ ,
\end{align}
where $\tilde r^2=\Phi(\tilde r) r^2$ and $\Phi(\tilde r)\equiv f'(\mathcal{R})=1+2\kappa c^2\beta \rho(\tilde r)$. After transforming back to the 
Jordan frame, taking the Taylor expansion around $\beta=0$ we may write down the modified hydrostatic equilibrium equation as
\begin{equation}\label{pres}
 p'=-g\rho(1+\kappa c^2 \beta [r\rho'-3\rho]) \ ,
\end{equation}
where $g=\text{const}$ is the surface gravity, which can be approximated on the star's surface (that is, on the photosphere, which 
is often taken as the surface of a star) as
\begin{equation}\label{surf}
 g\equiv\frac{G m(r)}{r^2}\sim\frac{GM}{R^2},
\end{equation}
where $M=m(R)$.
Let us notice that the transformation of the mass function $ m(\tilde r)$ to $m(r)$ depends on the energy density which on the
 non-relativistic star's surface will drop to zero. Due to that fact, we approximate the mass function to the one of the 
 very familiar form $ m'(r)=4\pi r^2\rho(r)$ in the Jordan frame, such that one has
 \begin{equation}
  m''=8\pi r\rho+4\pi r^2 \rho'.
\end{equation}
We use it in \ref{pres}, together with \ref{surf} written after differentation with respect to $r$ as 
$d^2m/dr^2=2m/r^2$, in order to find the following form
\begin{equation}\label{hyd}
 p'=-g\rho\left( 1+8\beta\frac{g}{c^2 r} \right).
\end{equation}

The heat transport with respect to radiative and conductive processes is given by \cite{hansen}
\begin{equation}\label{heat}
 \frac{\partial T}{\partial  m}=-\frac{3}{64\pi^2ac}\frac{\kappa_{rc}l}{r^4T^3},
\end{equation}
where $l$ is the local luminosity, the radiation density constant is $a=7.57\times 10^{-15}\frac{erg}{cm^3K^4}$ and the opacity 
$\kappa_{rc}$ is given by
\begin{equation}
 \frac{1}{\kappa_{rc}}=\frac{1}{\kappa_{rad}}+\frac{1}{\kappa_{cd}}
\end{equation}
with $\kappa_{rad}$ being the radiative opacity while $\kappa_{cd}$ the conductive one. Writing \ref{hyd} as 
\begin{equation}
 \frac{\partial p}{\partial m}=-\frac{G m}{4\pi r^4}\left(1+8\beta\frac{G m}{c^2 r^3}\right)
\end{equation}
and using it together with the heat transport \ref{heat} one has
\begin{equation}
 \frac{\partial T}{\partial p}=\frac{3\kappa_{rc}l}{16\pi acG mT^3}\left(1+8\beta\frac{G m}{c^2 r^3}\right)^{-1}.
\end{equation}
Similarly as in the standard case, we define a gradient describing the temperature variation with 
depth 
\begin{equation}
 \nabla_{\text{rad}}:=\left(\frac{d \ln{T}}{d\ln{p}}\right)_{\text{rad}}
\end{equation}
which in Palatini case takes a form
\begin{equation}\label{grad}
  \nabla_{\text{rad}}=\frac{3\kappa_{rc}lp}{16\pi acG mT^4}\left(1+8\beta\frac{G m}{c^2 r^3}\right)^{-1}.
\end{equation}

\subsection{Polytropic Palatini stars}\label{Sec_pol}

Since in the further part we will consider polytropic stars whose equation of state is given by the simple power-law relation
\begin{equation}\label{pol}
  p=K\rho^\gamma,
\end{equation}
it is convenient to recall now the Palatini Lane-Emden equation \cite{aneta2}. 
Its solutions will be needed to the discussion on the pre-main-sequence phase of the stellar evolution, as well as to describe fully convective stars on the Main 
Sequence. Thus, 
the modified Lane-Emden equation has the following form
\begin{equation}\label{LE}
 \frac{1}{\xi}\frac{d^2}{d\xi^2}\left[\sqrt{\Phi}\xi\left(\theta-\frac{2\kappa^2 c^2\rho_c\alpha}{n+1}\theta^{n+1}\right)\right]=
 -\frac{(\Phi+\frac{1}{2}\xi\frac{d\Phi}{d\xi})^2}{\sqrt{\Phi}}\theta^n,
\end{equation}
where $\Phi=1+2\alpha \theta^n$ with $\alpha$ defined as $\alpha=\kappa c^2\beta\rho_c$, while the dimensionless variables are given by
\begin{align}
 r&=r_c\xi,\;\;\;\rho=\rho_c\theta^n,\;\;\;p=p_c\theta^{n+1},\label{def1}\\
 r^2_c&=\frac{(n+1)p_c}{4\pi G\rho^2_c},\label{def2}
\end{align}
where $p_c$ and $\rho_c$ are the central pressures and densities and $n=\frac{1}{\gamma-1}$ is the polytropic index of the polytropic 
equation of state \ref{pol}. Let us notice that 
in the standard version of the Lane-Emden equation (when $\alpha=0$) one deals with only one parameter $n$ whose value is related to a type of the star. It also indicates the stable stars' configurations (see for example \cite{glen}). The most important values are $n=3/2$, modelling cores of fully convective stars and small white dwarfs, and $n=3$, used for high mass white dwarfs and approximated analysis of main-sequence stars. In a modified version we have the extra parameter $\alpha$, coming from the modification of GR, used for constraining the theory against observational data (or in some cases, by theoretical analysis, as for instance the mentioned stability problem, discussed in \cite{aneta3}).

The equation \ref{LE} has two exact solutions \cite{artur} for $n=\{0,1\}$
\begin{equation}
 \theta_{n=0}=-\frac{\xi^2}{6}+1,\;\;\;\;\;\theta_{n=1}=\frac{\xi^2-15}{2\alpha\kappa^2 c^2\rho_c(10+\xi^2)},
\end{equation}
thus for the other values of the index $n$ one needs to solve the equation numerically \cite{aneta2,gonzalo}. Let us notice that the equation depends 
on the central energy density which is a common feature of Palatini theories of gravity. That is, the theory
introduces new energy-density dependent contributions which distinguishes it from other proposals extending GR \cite{gonzalo2}.

Using the solutions of the modified Lane-Emden equation \ref{LE} one may obtain 
the star's mass, radius, central density, and temperature via the well-known expressions
\begin{align}
 M&=4\pi r_c^3\rho_c\omega_n,\\
 R&=\gamma_n\left(\frac{K}{G}\right)^\frac{n}{3-n}M^\frac{n-1}{n-3} \label{radiuss},\\
 \rho_c&=\delta_n\left(\frac{3M}{4\pi R^3}\right) \label{rho0s} ,\\
 T&=\frac{K\mu}{k_B}\rho_c^\frac{1}{n}\theta_n \label{temps},
\end{align}
where $k_B$ is the Boltzmann constant and $\mu$ the mean molecular weight. It should be commented that the constants \ref{omega} and \ref{delta}
\begin{align}
 \omega_n&=-\frac{\xi^2\Phi^\frac{3}{2}}{1+\frac{1}{2}\xi\frac{\Phi_\xi}{\Phi}}\frac{d\theta}{d\xi}\mid_{\xi=\xi_R},\label{omega}\\
  \gamma_n&=(4\pi)^\frac{1}{n-3}(n+1)^\frac{n}{3-n}\omega_n^\frac{n-1}{3-n}\xi_R,\label{gamma}\\
 \delta_n&=-\frac{\xi_R}{3\frac{\Phi^{-\frac{1}{2}}}{1+\frac{1}{2}\xi\frac{\Phi_\xi}{\Phi}}\frac{d\theta}{d\xi}\mid_{\xi=\xi_R}} \ . \label{delta}
\end{align}
 differ from their GR forms because of the new $\Phi$-dependent terms \cite{artur}. It is so since in order to obtain the equation
 \ref{LE}, the Einstein frame was used and finally one has to come back to the Jordan one by performing the conformal transformation.
 
Let us notice that in the case of polytropies the equation \ref{grad} can be written in terms of solutions of the 
modified Lane-Emden equation \cite{aneta2, gonzalo}
\begin{equation}
 \nabla_{\text{rad}}=\frac{3\kappa_{rc}lp}{16\pi acG mT^4}\left(1-\frac{4\alpha}{3\delta_n}\right)^{-1},
\end{equation}
with $\alpha=\kappa c^2\beta\rho_c$.
 
\section{A toy model for fully convective stars in Palatini gravity}
\subsection{A brief comment on dynamical instability}\label{stab}

Apart from the radiative and conductive energy transport briefly mentioned in \ref{NRs}, convection is another phenomenon which may have an important role in the heat
 transport in some regions of the star. It appears when 
small fluctuations of functions and variables describing a spherical symmetric star, which are always present in the star's interior, 
grow: that causes mixing of the stellar material as well as it may be an agent of energy transport through one region to another.

We will focus on an ideal gas, therefore $\rho\sim p/T$. Let us consider an element $\rho_e$ which remains always in the
pressure balance with the surrounding $\rho_s$, so $Dp:=p_e-p_s=0$. When the element is slightly hotter with $DT>0$, from the ideal gas relation we
have $D\rho<0$ - the element is lighter than the surrounding material and will be lifted upwards by the buoyancy forces from $r$ to $r+\Delta r$. 
The change of the element density risen by $dr$ is written as
\begin{equation}\label{stab1}
 D\rho=\left[\left(\frac{d\rho}{dr}\right)_e-\left(\frac{d\rho}{dr}\right)_s\right]\Delta r.
\end{equation}
If on the considered layer we deal with $D\rho>0$, the element is heavier than the surrounding and will be drawn back to its original position, so the perturbation 
is removed and we deal with a stable configuration.

Assuming that the energy is not being exchanged during that process (that is, the element rises adiabatically), we may rewrite the stability
condition \ref{stab1} for the equation of state $\rho=\rho(p,T,\mu)$ with the homogeneous chemical composition $\mu$ (which results as
 $d\mu=0$ for both, the element and surrounding) in the following form
\begin{equation}
 -\left(\frac{1}{T}\frac{dT}{dr}\right)_e+\left(\frac{1}{T}\frac{dT}{dr}\right)_s>0.
\end{equation}
Multiplying it by the term 
$-p\frac{dr}{dp}$ one obtains the stability criterion
\begin{equation}
 \left(\frac{d \ln{T}}{d\ln{p}}\right)_s<\left(\frac{d \ln{T}}{d\ln{p}}\right)_e,\;\;\text{or}\;\;\nabla<\nabla_e.
\end{equation}
If the element changes adiabatically, we may write $\nabla_e=\nabla_{\text{ad}}$ while if the energy is transported by the radiation (and conduction), 
then $\nabla=\nabla_\text{rad}$. This is the Schwarzschild criterion for the stable star's layer:
\begin{equation}\label{warun}
 \nabla_\text{rad}<\nabla_\text{ad}.
\end{equation}
When perfect, monatomic gas is considered, then the adiabatic temperature gradient can be shown to be $\nabla_\text{ad}$=0.4 (see e.g. \cite{ewol}).

However, if $\nabla_\text{rad}$ is too high, that is, we are dealing with large flux $F=l/(4\pi r^2)$ or very opaque matter, or $\nabla_\text{rad}$ has a
depression, the LHS of \ref{warun} will be bigger than $\nabla_\text{ad}$ and a part of the flux will be carried by the 
convection, so $\nabla\neq\nabla_\text{rad}$. Thus, the condition for the convective energy transport in some region of 
the star is $ \nabla_\text{rad}>\nabla_\text{ad}$.

Let us just comment that the chemical composition gradient $\nabla_\mu$, which will appear on the RHS when non-homogeneous chemical composition is considered (Ledoux criterion), has 
a stabilizing effect.

Since it was shown that in Palatini gravity the radiative gradient is modified, it will also have an effect on the Schwarzschild criterion, that is,
the convection appears when 
\begin{equation}
 \frac{3\kappa_{rc}pl}{16\pi acG mT^4}\left(1+8\beta\frac{G m}{c^2 r^3}\right)^{-1}>\nabla_\text{ad}.
\end{equation}
Therefore, depending on the Starobinsky parameter $\beta$, the modification can have a stabilizing or destabilizing effect.


\subsection{Convective stars}
\subsubsection{Hayashi tracks}\label{haytra}
We will consider a fully ionized monatomic gas with the temperature $T$ and mean molecular weight $\mu$ fulling the interior of a convective star up 
to the photosphere. We assume that the photosphere lies in $r\sim R$, where $R$ is the star's radius, and thus, as already mentioned, the 
stratification $\nabla_e=d\ln{T}/d\ln{p}=\nabla_\text{ad}$ is adiabatic and equaled to $2/5$. In such a case it turns out that the equation of state can be written as the polytropic 
equation of state \ref{pol} with the index $n=3/2$. Using the ideal gas relation in the polytropic EoS \ref{pol}
\begin{equation}
 \rho=\frac{\mu p}{N_A k_B T},
\end{equation}
where $N_A$ and $k_B$ are the Avogardo and Boltzmann constants, respectively, one may write
\begin{equation}\label{eos}
 p=\tilde K T^{1+n},\;\;\;\tilde K=\left(\frac{N_Ak_B}{\mu}\right)^{1+n}K^{-n}.
\end{equation}
Let us notice that however $K$ is a constant, it depends on modified gravity, since the formula
\begin{equation}\label{ka}
 K=\left[\frac{4\pi}{\xi_R^{n+1}(-\theta'_n(\xi_R))^{n-1}}\right]^\frac{1}{n}\frac{G}{n+1}M^{1-\frac{1}{n}}R^{\frac{3}{n}-1}
\end{equation}
includes the solutions of the modified Lane-Emden equation \ref{LE}. So $K$ and $\tilde K$ vary not only from star to star, but it also depends on the 
modified gravity model (here via the solutions of the modified Lane-Emden equation for given Starobinsky parameter $\beta$).

The EoS \ref{eos} holds as far as the photosphere. The photosphere is a visible surface with the temperature $T_\text{eff}$ which satisfies the Stefan-Boltzmann 
equation (that is, photosphere is a surface from which the radiation is emitted into space while $T_\text{eff}$ is the temperature of a black body). For the photosphere, 
the optical depth $\tau$ takes the value $\tau=2/3$. Above the photosphere one deals with an atmosphere that we assume to be radiative for which the 
absorption law is given by a simple relation
\begin{equation}\label{abs}
 \kappa_\text{abs}= \kappa_0 p^i T^j.
\end{equation}
Since we will consider rather cool stars, for temperatures in the range $3000\lesssim T\lesssim6000$K, its surface layer
is dominated by H$^{-}$ opacity \cite{hansen}, which for hydrogen mass fraction 
$X\approx0.7$ is
\begin{equation}\label{hydro}
 \kappa_{H^-}=\kappa_0 \rho^\frac{1}{2}\,T^9\,\,\text{cm}^2\text{g}^{-1},
\end{equation}
where $\kappa_0\approx2.5\times10^{-31} \left(\frac{Z}{0.02}\right)$. A metal mass fraction $Z$ is
taken from the range $0.001\lesssim Z\lesssim0.03$, with $0.02$ being the solar metallicity. For the ideal gas,  
\ref{hydro} can be rewritten as
\begin{equation}\label{hydro2}
 \kappa_{H^-}=\kappa_g p^\frac{1}{2}\, T^{8.5}\,\,\text{cm}^2\text{g}^{-1},
\end{equation}
where $\kappa_g=\kappa_0\left(\frac{\mu}{N_Ak_B}\right)^\frac{1}{2}\approx1.371\times10^{-33}Z\mu^\frac{1}{2}$. 

As already mentioned, the photosphere can be defined at the radius for which the optical depth with a mean opacity $\kappa$ (averaged over the stellar 
atmosphere) is equaled to $2/3$:
\begin{equation} \label{eq:od}
 \tau(r)=\kappa\int_r^\infty \rho dr=\frac{2}{3}.
\end{equation}
Using this relation in order to integrate the hydrostatic equilibrium equation \ref{hyd} with $r=R$ and $M=m(R)$, and applying the absorption law \ref{hydro2} one gets
\begin{equation}\label{fotos}
 p_\text{ph}=8.12\times10^{14}\left(\frac{ M\left(1-\frac{4\alpha}{3\delta}\right)}{L T_\text{ph}^{4.5}Z\mu^\frac{1}{2}}\right)^\frac{2}{3},
\end{equation}
where we have already used the Stefan-Boltzmann law $L=4\pi\sigma R^2T^4_\text{ph}$ with $ T_{\text{eff}{\mid_{r=R}}}\equiv T_\text{ph}$.

On the other hand, from \ref{eos} taken on the photosphere with $n=3/2$ we have
\begin{equation}
 T_{\text{eff}{\mid_{r=R}}}=\left(\frac{\mu}{N_Ak_B}\right)^{-2/3}\left(\frac{4\pi}{\xi (-\theta')^\frac{1}{2}}\right)^{2/5}
 \left(\frac{2G}{5}\right)^\frac{3}{5}M^\frac{1}{5}R^\frac{3}{5}p_\text{ph}^\frac{2}{5}.
\end{equation}
Applying again the Stefan-Boltzmann law to the above expression, one gets 
\begin{equation}
  T_\text{ph}=9.196\times10^{-6}\left( \frac{L^\frac{3}{2}Mp_\text{ph}^2\mu^5}{-\theta'\xi_R^5} \right)^\frac{1}{11}.
\end{equation}
The pressure appearing in the above relation must be equaled to the gravitational pressure taken on the photosphere, that is, the equation 
\ref{fotos}. After some algebraic manipulation and rescaling the result to the solar values $L\rightarrow L/L_\odot$ and $M\rightarrow M/M_\odot$, where $L_\odot$ and 
$M_\odot$ are solar luminosity and solar mass, respectively, one finds
\begin{equation}\label{hay}
 T_\text{ph}=2487.77\mu^\frac{13}{51}\left( \frac{L}{L_\odot}  \right)^{\frac{1}{102}}\left( \frac{M}{M_\odot}  \right)^{\frac{7}{51}}
 \left( \frac{\left(\frac{1-\frac{4\alpha}{3\delta}}{Z}\right)^\frac{4}{3}}{\xi_R^5\sqrt{-\theta'}} \right)^\frac{1}{17}\text{K}.
\end{equation}

Treating the star's mass and mean molecular weight as parameters (so each star with a mass $M$ and uniform composition $\mu$ will have 
its own evolutionary track drawn by \ref{hay}), and for given metallicity $Z$, we got a familiar form relating the effective temperature and luminosity of the pre-main-sequence
star.
The tracks given by the equation \ref{hay} are Hayashi tracks \cite{hayashi}, that is, almost vertical lines on the right-hand side of the H-R diagram. 
They are followed by the infant stars with masses not exceeding three solar masses, having nearly constant low effective temperature. The stars on the Hayashi tracks are fully convective 
apart from the radiative photosphere. The relation \ref{hay} shows that their effective temperatures are almost independent of luminosity which means that $T_\text{eff}$ is nearly independent 
of the way how the energy is generated. 

\begin{figure}[t]
\centering
\includegraphics[scale=.60]{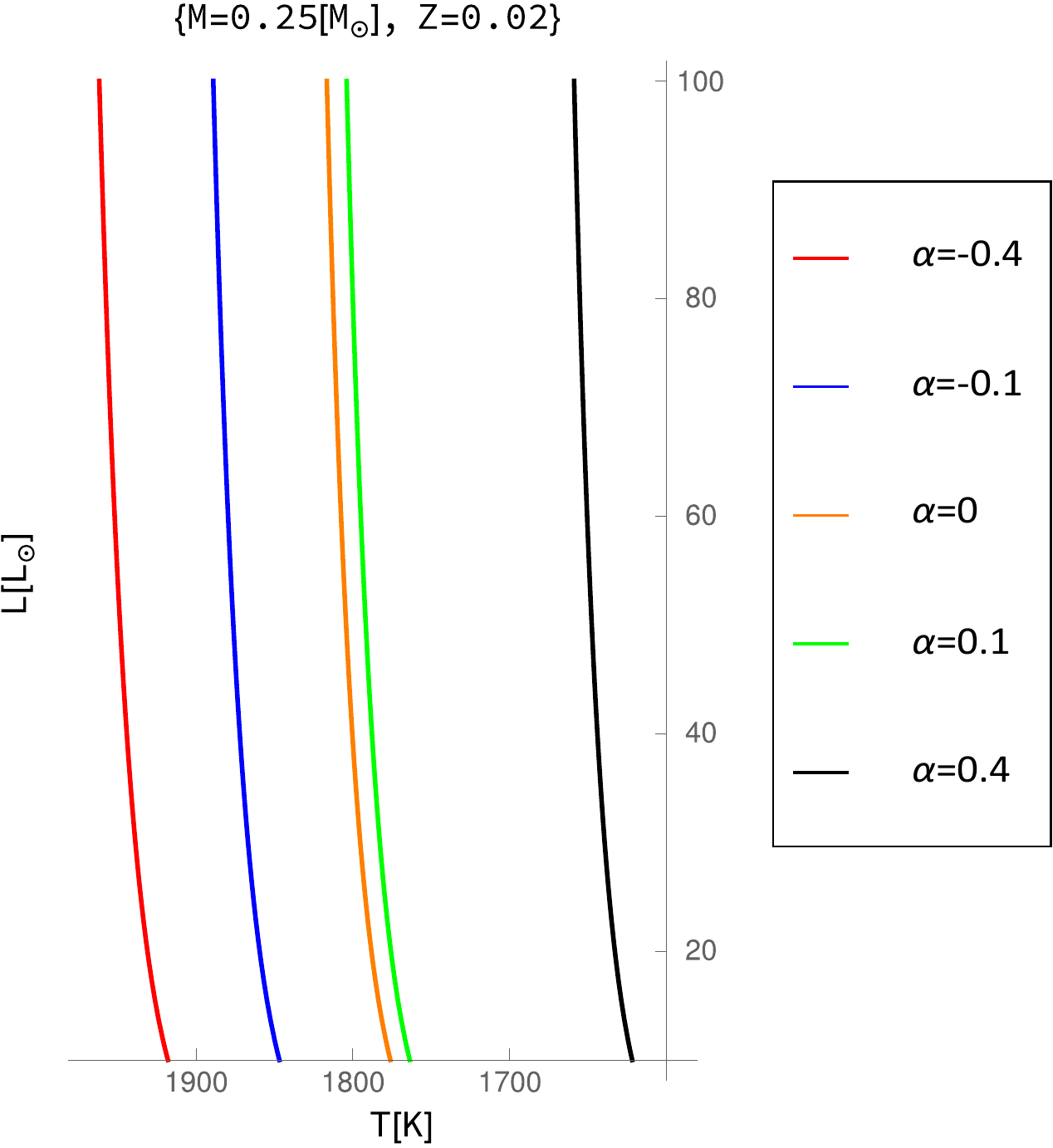}
\caption{[color online] The Hayashi tracks of a star with mass $M=0.25M_\odot$, metallicity $Z=0.02$, and chemical composition $\mu=0.618$ with respect to
a few values of the parameter $\alpha$, given by the equation \ref{hay}.}
\label{fig.1}
\end{figure}

 Let us notice that the temperature coefficient is too low (it should be around $4000$K) but this is the result of a very simplified calculation.
However, as already mentioned, the considered toy-model is good enough to examine the modified gravity effects. 
A few Hayashi tracks with respect to the parameter $\alpha$ are drawn in 
the Figure \ref{fig.1} for a star with mass $M=0.25M_\odot$, mean molecular weight $\mu=0.618$, and solar metallicity $Z=0.02$. Bigger masses 
and smaller amounts of metals 
($Z<0.02$) give higher effective temperatures with a similar pattern as in Figure \ref{fig.1} for different parameters $\alpha$.

 Coming back to the relation \ref{hay}, let us emphasize again that the difference between the GR relation and ours is not only given by the presence of the 
parameter $\alpha$ but also by the values of $\xi_R$ and $\theta'$, which are different in modified gravity - they are obtained for a given value of $\alpha$ by 
solving the modified Lane-Emden equation \ref{LE}.
Even a small change in the constants, which will be our case, changes the position of the Hayashi track on the H-R diagram. Therefore,
for a non-zero Starobinsky parameter we will deal with a shift of such a path which immediately leads to the possibility to constrain the theory 
by the observations of fully convective pre-main sequence stars following Hayashi tracks. 
Especially useful for such analysis could be T Tauri stars (for the review, see for example \cite{bert}) which are newly formed low-mass stars, very active and 
variable, in the process of contracting to the Main Sequence, and which just started to
be visible in the optical range.

\subsubsection{Fully convective stars on Main Sequence}
Let us now consider a fully convective star (apart from the radiative region on the surface) at the end of its journey on 
the Hayashi track, that is, a PMS star approaching the Main Sequence. Such an object, depending on its mass, may become a Brown Dwarf \cite{Burrows:1992fg}, convective star on 
the Main Sequence or, before reaching the Main Sequence, it may start following the Henyey track as soon as the radiative core appears \cite{hll,hbg,hvb}. The last evolutionary track is 
represented by the almost horizontal lines on the H-R diagram (their luminosities remain almost constant while their effective temperatures increase). 

When a star on the Hayashi track contracts, its luminosity 
decreases and thus it may happen that the star's interior becomes radiative (the turning point from Hayashi to Henyey track). However, the case when fully 
convective star reaches the Main Sequence before developing a radiative core also occurs, and this is the situation which we would like to study in our toy model.

As discussed in \ref{stab}, a chemically homogeneous layer is convective if 
\begin{equation}
 \nabla_\text{rad}>\nabla_\text{ad}.
\end{equation}
A radiative region in the star's center starts to develop when $ \nabla_\text{rad}$ drops below $\nabla_\text{ad}$; thus let us
examine the limiting case, that is, when $ \nabla_\text{rad}=\nabla_\text{ad}$.
Therefore, for the polytropic model \ref{eos} with the polytropic index $n=\frac{3}{2}$ and the assumption that the deep interior opacity can be described by the Kramers' absorption law \ref{abs} with $i=1$ and $j=-4.5$, 
the temperature gradient $\nabla_\text{rad}$ after entering the numerical values for the constants is
\begin{equation}
 \nabla_\text{rad}=1.564\times10^{70}\frac{\delta_{3/2}\xi^{5}\theta' }{(4\alpha-3\delta_{3/2})}\frac{\kappa_0L}{\mu^5 M^2 R^3 T^{3.5}},
\end{equation}
where $L$ is the total luminosity (after the homology law of contracting stars \cite{szyd}).
Applying the central temperature solution \ref{temps} together with the homology contraction argument
\begin{equation}
 T_c=6.679\times10^{-16}\frac{\mu\delta_{3/2}^{2/3}}{\xi^{5/3}(-\theta')^{1/3}}\frac{M}{R},
\end{equation}
and the Stefan-Boltzmann law one gets
\begin{equation}
 \nabla_\text{rad}=2.71\times10^{-12} \frac{\xi^{10.83}(-\theta')^{2.167}}{\delta_{3/2}^{1.33}(3\delta_{3/2}-4\alpha)}  
 \frac{\kappa_0\left(\frac{L}{L_\odot}\right)^{1.25}}{\mu^{8.5}M_{-1}^{5.5}T_\text{eff}},
\end{equation}
where we have defined $M_{-1}=M/(0.1M_\odot)$.
For the ideal gas model that we are using the adiabatic gradient is $\nabla_\text{ad}=0.4$, thus the minimum luminosity 
\begin{equation}\label{lmin}
 L_\text{min}=9.89\times10^{7}L_\odot  \frac{\delta_{3/2}^{1.064}(\frac{3}{4}\delta_{3/2}-\alpha)}{\xi^{8.67}(-\theta')^{1.73}} \left(\frac{T_\text{eff}}{\kappa_0}\right)^{0.8} M_{-1}^{4.4},
\end{equation}
 which is, for the GR values in the case of a star with $T_\text{eff}=4000$ K, mass $M=0.1M_\odot$ and opacity \ref{ff}, around $L^\text{GR}_\text{min}=0.66\times10^{-4}L_\odot$ where $L_\odot\approx 4\times 10^{33}\text{ergs/s}$. Then, we may write for the arbitrary low-mass star
\begin{equation}
 L^\text{mod}_\text{min}\approx   181.8 L^\text{GR}_\text{min}  \frac{\delta_{3/2}^{1.064}(\frac{3}{4}\delta_{3/2}-\alpha)}{\xi^{8.67}(-\theta')^{1.73}}.
\end{equation}

It may however happen that a low-mass star moving along the Hayashi track can cross the Main Sequence without reaching the minimum luminosity \ref{lmin} needed for developping 
a radiative region in its center. It means that such a star may reach the Main Sequence being fully convective (apart from the photosphere), or, if its mass does not 
exceed the value $\sim 0.09M_\odot$, it will fail to be a Main Sequence star, that is, the star will become a Brown Dwarf
(\cite{kumar}, for the review see \cite{Burrows:1992fg},
in modified gravity \cite{gonzalo,sak1,sak2,cris}).

Let us assume that such a fully convective star has smaller luminosity but very close to the one obtained above (so it almost has the
luminosity $L_\text {min}$), and that its mass is large enough to burn hydrogen. It means that we have a fully convective star on the Main Sequence, while another star with the luminosity (and mass) a bit bigger than $L_\text{min}$ will have a radiative center. For both stars the energy generation per unit mass in the process of hydrogen burning can be obtained by the
following power-law form \cite{Burrows:1992fg}
\begin{equation} \label{eq:pp}
\dot{\epsilon}_{pp}= \dot{\epsilon}_c \left(\frac{T}{T_c}\right)^s \left(\frac{\rho}{\rho_c} \right)^{u-1} \ ,
\end{equation}
with $s \approx 6.31$ and $u \approx 2.28$, while $T_c$ and $\rho_c$ are as usual the central temperature and density, respectively, obtained from the near 
center solution of \ref{LE}.
As before, the hydrogen fraction is taken as $X\approx0.75$, while $\dot{\epsilon}_c\approx 3.4\times10^{-9} T_c^s\rho_c^{u-1}$
ergs g$^{-1}$s$^{-1}$ \cite{Burrows:1992fg}. Integrating it over 
the stellar volume one gets the luminosity from hydrogen burning
 \begin{equation}
  L_{HB}=4\pi r_c^3\rho_c\dot{\epsilon}_c\int^{\xi_R}_{0}\xi^2\theta^{n(u+\frac{2}{3}s)}d\xi,
 \end{equation}
 where $\theta$ is a solution of the generalized Lane-Emden equation.
Following the result found in \cite{gonzalo}, we have for the quadratic Palatini case that the luminosity from hydrogen burning is given by the following 
expression
\begin{equation}\label{lhb}
 L_\text{HB}=1.53\times10^7L_\odot\frac{\delta^{5.487}_{3/2}}{\omega_{3/2}\gamma^{16.46}_{3/2}}M^{11.977}_{-1}
 \frac{\eta^{10.15}}{(\eta+\alpha_d)^{16.46}},
\end{equation}
where
$\eta$ measures the degree of the degeneracy electron pressure of the star and $\alpha_d\equiv\frac{5\mu_e}{2\mu}\approx4.82$. Let us notice again that 
the above relation is modified by the values of $\delta$, $\omega$, and $\gamma$. 
 Let us just mention that the above luminosity for GR values for a star with mass $M=0.1 M_\odot$ is 
around $L_\text{HB}=0.61\times10^{-4}L_\odot$ \cite{Burrows:1992fg, gonzalo}, being thus a very faint red dwarf star which belongs to the class M. It can be also written for a general low-mass star as
\begin{equation}
 L^\text{mod}_\text{HB}\approx   198 L^\text{GR}_\text{HB}  \frac{\delta^{5.487}_{3/2}}{\omega_{3/2}\gamma^{16.46}_{3/2}}.
\end{equation}

Equaling the two luminosities \ref{lmin} and \ref{lhb} allows to find the mass of the biggest fully convective star on 
the Main Sequence (a star onset of radiative core development)
\begin{equation}\label{masss}
 M_{-1}=1.7\frac{\mu^{0.9}T_\text{eff}^{0.11}(\alpha_d+\eta)^{2.173}}{\eta^{1.34}\kappa_0^{0.11}}
 \frac{\gamma^{2.173}\omega^{0.132}}{\delta_{3/2}^{0.58}\xi^{1.14}(-\theta')^{0.23}}.
\end{equation}

The obtained expression, as we will see below, is sensitive to opacity (here to Kramers' opacity with $\kappa_0$). Moreover, one usually solves much more complicated equations numerically, having tabulated values of opacities and 
taking into account nuclear reaction rates, more accurate EoS, non-grey atmosphere models, ...\cite{chab, cant}). From such a theoretical analysis it is well-known that the fully convective stars on the Main Sequence have masses from the range $(\sim0.09-0.35)M_\odot$. The early analysis gave 
the upper bound $0.26M_\odot$ \cite{haynak} while $0.35M_\odot$ was obtained by \cite{chab}. In our very simplified model one gets masses 
a bit above the minimum one needed for hydrogen burning. However, as argumented above, we may use this theoretical crude result to see the modified gravity influence on the stars' masses.

Using the GR values for the Lane-Emden solutions ($\alpha=0$), as well as $\alpha_d=4.82$ and the degree of the degeneracy electron pressure
as $\eta=9.4$ \cite{Burrows:1992fg} one gets
\begin{equation}
 M_{-1}^\text{GR}=31.17 \mu^{0.9} T_\text{eff} \kappa_0^{-0.11},
\end{equation}
while taking the mean molecular weight $\mu=0.618$ and the effective temperature as $T_\text{eff}=4000$K:
\begin{equation}
 M=4.86M_\odot\kappa^{-0.11}_0.
\end{equation}
Let us consider two estimated opacities of the Kramers' form \ref{abs} with $i=1,\; j=-4.5$: the total bound-free and free-free opacities \cite{hansen}:
\begin{align}
 \kappa_0^\text{bf}&\approx 4\times10^{25}\mu \frac{ Z(1+X)}{N_Ak_B}\text{cm}^2\text{g}^{-1},\label{ff}\\
  \kappa_0^\text{ff}&\approx 4\times10^{22}\mu\frac{(X+Y)(1+X)}{N_Ak_B}\text{cm}^2\text{g}^{-1},
\end{align}
for which the masses are, taking $X=0.75$ and $Z=0.02$, respectively
\begin{align}
 M_\text{bf}=0.099 M_\odot,\;\;\;M_\text{ff}=0.135 M_\odot.
\end{align}
As already mentioned, from our very simplified model we got the mass values too small. Even so, we use these values as the reference ones to see how much they are affected by modification of our gravitational model. The results are presented in
the Table \ref{tab} for which we have calculated a few masses for both opacity models with respect to the 
parameter $\alpha$ (and the corresponding quantities depending on it, such as $\omega$, $\gamma$, $\delta$ \ref{omega}).

\section{Conclusions}
We would like to stress again that the presented model is too simple to describe real PMS stars and fully convective M dwarfs on the Main Sequence. We have not taken into account the magnetic field \cite{cant,morin}, thermonuclear depletion of the light elements \cite{Bildsten,ush,aneta4}, rotation \cite{wright}, protostellar initial conditions, more accurate atmosphere and opacities models.
However, our aim was to demonstrate how modified gravity can affect the macroscopic values of such objects, such as mass (which is the most crucial 
quantity in order to study stars' evolution), radius,
effective temperature and luminosity as well as how it may change their early evolutionary tracks. Due to that fact, 
the considered model is close enough to young low-mass stellar objects in order to show such an effect.

The discussion undertaken in the subsection \ref{haytra} proved that the Hayashi tracks in modified gravity are significantly shifted to right or left in the H-R diagram relative to their GR counterpart. In the GR case (when $\alpha=0$), the equation \ref{hay} describes a path of effective 
temperature which is taken by a PMS star with the mass $M$ and mean molecular weight $\mu$ - at any stage the temperature cannot fall below that value because the star could 
enter into the forbidden region with lower temperatures on the H-R diagram \cite{hayashi}. Since in Palatini gravity this value is modified, the PMS stars with known mass/luminosity moving along the Hayashi tracks could be used to constraint the theory \cite{hill,tout}.
 By considering more realistic models which take into account missing physics (for example proper treating of the atmosphere, opacity, degeneracy, rotation,...), analysis of Hayashi tracks of the PMS, such as T-Tauri stars \cite{bert},
will be a powerful tool to constrain gravitational theories which modify the stellar equations.

It should be also noticed that for a star with mass $M$, metallicity $Z$, and uniform composition $\mu$ the effective temperature
can be higher (lower) because of the extra term coming from the modification of the gravitational theory (in other words, the star 
would follow a neighbor's track of the one given by GR). Having a slightly different evolutionary track means that the star may stay longer or shorter on the pre-main sequence phase which has an effect on the total stars' luminosity contributing to 
the galaxy brightness (a very nice discussion on that topic can be found in \cite{davis}).

Another important point which also appeared in our consideration is the direct effect of modified gravity on the equation of state. It was already pointed out in 
\cite{kim} (therein EiBI theory \cite{vol,banados,delsate2,BeltranJimenez:2017doy} studied) and later in \cite{adria} (general metric-affine theories) that the modified gravity introduces gravitational backreaction on the fluid particles, resulting as an additional pressure which should be taken into account. Commonly used EoSs, such as polytropic ones, are obtained by considering motion of particles in flat spacetime so it may happen that the pressure may not be local in high curvature regime which means that the covariant assumption for the pressure might not hold \cite{reva}. In our simple model we have
used a modified polytropic EoS: the polytropic constant $K$ \ref{ka} turn out to depend on solutions of generalized Lane-Emden equation and thus 
is sensitive to modifications of gravitational equations. 

A similar situation happens with the temperature gradient which is used for examinations of dynamical stability. As demonstrated, the Schwarzschild (or Ledoux) criterion is also altered by the theory of gravity in such a way that the extra term appearing in the criterion has stabilizing or destabilizing effect. Depending on the sign of the parameter, the radiative core can develop quicker or slower (it means, for smaller or bigger masses) than for the masses predicted by GR \cite{chab}. In consequence, improving our toy model and then confronting it against accurate empirical masses could be a powerful tool to constraint the modified gravity.
Such empirical mass-luminosity relations are already available (see e.g. \cite{delf,mann1,pars,mann2}) as well as more direct and model-independent approaches for mass determination using detached eclipsing binaries (for a review see \cite{chab2,ser}). 

The accurate mass determination is crucial when one shapes the evolutionary track of a star - we have shown that the minimum mass 
a star needs to have in order to develop a radiative core can be different \ref{masss} than the one given by theoretical models using Newtonian 
hydrostatic equilibrium. When low-mass stars considered, fully convective stars and stars with radiative core are modelled in different ways, thus 
knowing that theoretical range of fully convective stars can have other 
upper limit (from $0.09$ to $0.35M_\odot$ in GR based theoretical models \cite{chab}) might improve existing numerical models. Moreover, that could also shed light on a discrepancy between predicted and dynamical masses 
of the M dwarfs and PMS stars with masses below $0.5M_\odot$, discussed in details in, for example, \cite{hill, siess}.

The general conclusion, together with the previous works on low-mass stars in modified gravity \cite{sak1,sak2, gonzalo, cris}, is that not only extreme environments such as 
black holes, neutron stars and early/late cosmology are a background for testing theories of gravity.
Low-mass stellar objects give additional, if not  simpler (taking into account still unknown internal features of neutron stars or conditions of early stage of the universe), possibility to have 
a closer look at a bunch of gravitational theories, for one deals with better understood density regimes allowing to examine eventual effects caused by such proposals.

\begin{table}[t!]
\begin{center}
\begin{tabular}{|c||c|c|}
\hline
$\alpha$ & $M_\text{bf}/M\odot$ & $M_\text{ff}/M\odot$  \\
\hline\hline
-0.4 &  0.047 & 0.065 \\ 
-0.1 & 0.083 & 0.114 \\ 
\hline
0 (GR)    & 0.099 & 0.135 \\ 
\hline
0.1  & 0.12  & 0.159 \\ 
0.4   & 0.18 & 0.24 \\  
\hline
\end{tabular}
\caption{Numerical values of maximal masses (in solar masses) of fully convective stars on the Main Sequence for different values 
of $\alpha=\kappa c^2 \beta \rho_c$. 
}
\label{tab}
\end{center}
\end{table}

\acknowledgments{The author would like to thank Jakub Ostrowski for enlightening discussions.
 
 The work is supported by the European Union through the ERDF CoE grant TK133.
 }


\begin{thebibliography}{99}


  \bibitem{Will:2014kxa}
  C.~M.~Will,
  Living Rev.\ Rel.\  {\bf 17} (2014) 4.

\bibitem{TheLIGOScientific:2017qsa}
  B.~P.~Abbott {\it et al.} [LIGO Scientific and Virgo Collaborations],
  Phys.\ Rev.\ Lett.\  {\bf 119} (2017)  161101.

\bibitem{Akiyama:2019cqa}
  K.~Akiyama {\it et al.} [Event Horizon Telescope Collaboration],  Astrophys.\ J.\  {\bf 875} (2019)  L1.

  \bibitem{Barack:2018yly}
  L.~Barack {\it et al.}, arXiv:1806.05195 [gr-qc].

\bibitem{Copeland:2006wr}
E. J. Copeland, M.~Sami, and S.~Tsujikawa, Int. J. Mod. Phys. D \textbf{15} (2006) 1753.

\bibitem{Nojiri:2006ri}
S.~Nojiri and S. D. Odintsov, Int. J. Geom. Meth. Mod. Phys. \textbf{4}  (2007)  115.

\bibitem{nojiri2} S. Nojiri, S.D. Odintsov, V.K. Oikonomou, {\it Modified Gravity Theories on a Nutshell: Inflation, Bounce and
Late-time Evolution}, Physics Reports 692 (2017).

\bibitem{nojiri3} S. Nojiri, S.D. Odintsov, {\it Unified cosmic history in modified gravity: from
$F(R)$ theory to Lorentz non-invariant models}, Physics Reports 505 (2011).

\bibitem{Capozziello:2007ec}
S.~Capozziello and M.~Francaviglia, Gen. Rel. Grav. \textbf{40} (2008) 357.

\bibitem{Carroll:2004de}
S. M. Carroll, A.~De~Felice, V.~Duvvuri, D. A. Easson, M.~Trodden, and M. S. Turner, Phys. Rev. D \textbf{71} (2005) 063513.

\bibitem{ParTom}
L. Parker and D. J. Toms, \emph{``Quantum Field Theory in Curved Spacetime: Quantized Fields and Gravity"}
(Cambridge University Press, Cambridge, England, 2009).

\bibitem{BirDav}
N. D. Birrel and P. C. W. Davies, \emph{``Quantum Fields in Curved Space"} (Cambridge University Press, Cambridge, England, 1982).

\bibitem{Senovilla:2014gza}
  J.~M.~M.~Senovilla and D.~Garfinkle,  Class.\ Quant.\ Grav.\  {\bf 32} (2015)  124008.
  
  \bibitem{junior} J.D. Toniato, R.C. Rodrigues, A. Wojnar, Phys.Rev.D 101 (2020) 6, 064050.
  
  \bibitem{sch} P.K. Schwartz, D. Giulini, Phys. Rev. A 100, 052116 (2019).
  
  \bibitem{ol1} G.J. Olmo, Phys. Rev.D77, 084021 (2008).
  
  \bibitem{ol2} G.J. Olmo, Phys. Rev. Lett. 98, 061101 (2007).

  \bibitem{DeFelice:2010aj}
A.~De~Felice and S.~Tsujikawa, Living Rev.\ Rel. \textbf{13} (2010) 3.

\bibitem{brans} C. H. Brans and R. H. Dicke, Phys. Rev. \textbf{124} (1961) 925.

\bibitem{Bergmann} P. G. Bergmann, Int. J. Theor. Phys.  \textbf{1} (1968) 25.

\bibitem{BeltranJimenez:2019tjy}
  J.~Beltr\'an Jim\'enez, L.~Heisenberg and T.~S.~Koivisto,
  arXiv:1903.06830 [hep-th].

\bibitem{Dabrowski:2012eb}  M. P.~Dabrowski and K.~Marosek, JCAP  \textbf{1302} (2013)  012.

\bibitem{Leszczynska:2014xba} K.~Leszczynska, A.~Balcerzak and M. P.~Dabrowski, JCAP \textbf{1502} (2015) 012.

  \bibitem{Ezquiaga:2017ekz}
  J.~M.~Ezquiaga and M.~Zumalac\'arregui,
Phys.\ Rev.\ Lett.\  {\bf 119} (2017) 251304.

\bibitem{Baker:2017hug}
  T.~Baker, E.~Bellini, P.~G.~Ferreira, M.~Lagos, J.~Noller and I.~Sawicki,
  Phys.\ Rev.\ Lett.\  {\bf 119} (2017) 251301.
  
  \bibitem{Creminelli:2017sry}
  P.~Creminelli and F.~Vernizzi,
  Phys.\ Rev.\ Lett.\  {\bf 119} (2017) 251302.
  
  \bibitem{Langlois:2017dyl}
  D.~Langlois, R.~Saito, D.~Yamauchi and K.~Noui,
  Phys.\ Rev.\ D {\bf 97} (2018) 061501.
  
  \bibitem{Sakstein:2017xjx}
  J.~Sakstein and B.~Jain,
  Phys.\ Rev.\ Lett.\  {\bf 119} (2017)  251303.
  
  \bibitem{Lombriser:2015sxa}
  L.~Lombriser and A.~Taylor,
  JCAP {\bf 1603} (2016) 031.

  
\bibitem{TheLIGOScientific:2017first}
  B.~P.~Abbott {\it et al.} [LIGO Scientific and Virgo Collaborations],
Phys. Rev. Lett. \textbf{116} (2016) 061102.
  
 \bibitem{Berti:2015itd}
  E.~Berti {\it et al.},  Class.\ Quant.\ Grav.\  {\bf 32} (2015) 243001.
  
  \bibitem{sun1} A. Casalino, M. Rinaldi, L. Sebastiani, S. Vagnozzi, Phys. Dark Univ. 22 (2018) 108
  
  \bibitem{sun2} A. Casalino, M. Rinaldi, L. Sebastiani, S. Vagnozzi, Class. Quant. Grav. 36 (2019) 017001

\bibitem{lina} M. Linares, T. Shahbaz, and J. Casares, The Astrophysical Journal \textbf{859} (2018) 54.

\bibitem{as} J. Antoniadis {\it et al.}, Science \textbf{340} (2012)  6131.

\bibitem{craw} F. Crawford, M. S. E. Roberts, J. W. T. Hessels, S. M. Ransom, M. Livingstone, C. R. Tam and V. M. Kaspi, Astrophys. J. \textbf{652} (2006) 1499.

\bibitem{NSBH} R. Abbott et al, The Astrophysical Journal 896 L44 (2020).

\bibitem{straight} M.C. Straight, J. Sakstein, E.J. Baxter, arXiv:2009.10716 .

 \bibitem{chab3} G. Chabrier. I. Baraffe, Annu. Rev. Astron. Astrophys.,  38: 337-77 (2000).
 
 \bibitem{lau} G. Laughlin, P. Bodenheimer, F.C. Adams, The Astrophysical Journal, 482: 420-432, 1997.
  
\bibitem{chab} G. Chabrier, I. Baraffe, Astronomy and Astrophysics, v.327, p.1039-1053 (1997).
     
     \bibitem{kumar} S.S. Kumar, The Astrophysical Journal 137 (1963): 1121.
     
     \bibitem{Burrows:1992fg}
  A.~Burrows and J.~Liebert,    Rev.\ Mod.\ Phys.\  {\bf 65} (1993) 301.
  
  \bibitem{bert} C. Bertout, {\it T Tauri stars: Wild as dust}, Anny. Rev. Astron. Astrophys. 27: 351-95, 1989.
 
 \bibitem{catelan} M. Catelan, AIP Conf.Proc. 930: 39-90, 2007.
 
 \bibitem{kroupa} P. Kroupa, Science, 295, 82 (2002).
 
 \bibitem{sand} A.R. Sandage, The Astronomical Journal 58 (1953): 61-75.
 
 \bibitem{cass} A. Cassan et al, Nature 481.7380 (2012): 167-169.
 
 \bibitem{ali} Y. Alibert, W. Benz, Astronomy and Astrophysics, 598, L5 (2017).
 
 \bibitem{harps1} M. Mayor, et al, Astronomy and Astrophysics 507.1 (2009): 487-494.
 
 \bibitem{harps2} X. Bonfils et al, Astronomy and Astrophysics 549 (2013): A109.
 
 \bibitem{yun} J. Yang, Y. Liu, Y. Hu, D.S. Abbot, The Astrophysical Journal, 796: L22, 2014.
 
 \bibitem{brad} B. Hansen, International Journal of Astrobiology, 14(2), 267-278 (2015).
 
 \bibitem{ramses} R.M. Ramirez, L. Kaltenegger, The Astronomical Journal 797: L25 (2014).
 
 \bibitem{bohn} A. Bohn et al, The Astrophysical Journal 898: L16 (2020).
 
 \bibitem{raf1} G.G. Raffelt, {\it Astrophysics probes of particle physics}, Physics Reports, Vol 333, 2000.
 
 \bibitem{raf2} G.G. Raffelt, {\it Stars as laboratories for fundamental physics: The astrophysics of neutrinos, 
 axions, and other weakly interacting particles}, University of Chicago press, 1996.
 
 \bibitem{vieira} J.P.P. Vieira, C.J.A.P. Martins, M.J.P.F.G. Monteiro, Phys.Rev.D 86 (2012) 043003.
 
  \bibitem{helio} I.D. Saltas, I. Lopes, Phys. Rev. Lett. 123 (2019) 9, 091103.
  
  \bibitem{kum} R.K. Jain, C. Kouvaris, N. G. Nielsen, Phys. Rev. Lett. 116, 151103.
 
  \bibitem{orf1} O. Bertolami and J. P\'{a}ramos, Phys. Rev. D 77, 084018.
 
  \bibitem{orf2} O. Bertolami and J. P\'{a}ramos, Phys. Rev. D 71, 023521. 
  
   \bibitem{orf3} O. Bertolami, J. PP\'{a}ramos, and P. Santos, Phys. Rev. D 80, 022001.
   
\bibitem{orf4} O. Bertolami, H. Mariji, Phys. Rev. D 93, 104046.
 
\bibitem{reva} G.J. Olmo, D. Rubiera-Gracia, A. Wojnar, {\it Stellar structure models in modified theories of gravity: lessons and challenges}, Physics Reports 876 (2020) 1-75.

\bibitem{bak} T. Baker et al {\it The Novel Probes Project--Tests of Gravity on Astrophysical Scales}, arXiv:1908.03430 (2019).

\bibitem{chang} P. Chang, L. Hui, The Astrophysical Journal 732.1 (2011): 25.

\bibitem{davis} A-Ch. Davis et al, Physical Review D 85.12 (2012): 123006.

\bibitem{chow} S. Chowdhury, T. Sarkar, arXiv:2008.12264.
 
  \bibitem{weinberg} S. Weinberg,  {\it Gravitation and Cosmology: principles and Applications of the General Theory of Relativity}, John Wiley and Sons (1972).
 
\bibitem{BSS} A. Stachowski, M. Szydlowski, A. Borowiec, Eur. Phys. J. C77, 406 (2017).

\bibitem{SSB} M. Szydlowski, A. Stachowski, A. Borowiec, Eur. Phys. J. C77, 603 (2017).

\bibitem{o} V.I. Afonso, G.J. Olmo, D. Rubiera-Garcia, Phys.Rev.D 97 (2018) 2, 021503.

\bibitem{o1} V.I. Afonso, G.J. Olmo, E. Orazi, D. Rubiera-Garcia, Eur.Phys.J.C 78 (2018) 10, 866.

\bibitem{o2} V.I. Afonso, G.J. Olmo, E. Orazi, D. Rubiera-Garcia, Phys.Rev.D 99 (2019) 4, 044040.

   \bibitem{aneta} A. Wojnar, Eur. Phys. J. C78 (2018) no.5, 421.

\bibitem{mr1} K. Kainulainen, V. Reijonen, D. Sunhede, Phys. Rev. D. 76 (2007) 043503.

\bibitem{mr2} F. A. T. Pannia, F. Garcia, S. E. P. Bergliaffa, M. Orellana, M., G. E. Romero, Gen. Rel. Grav. 49 (2017) 25.

\bibitem{mr3} D. E. Barraco, V. H. Hamity, Phys. Rev. D 62 (2000) 044027.

\bibitem{mr4} V. Reijonen, [arXiv:0912.0825 [gr-qc]].

\bibitem{mr5} G. Panotopoulos, Gen. Rel. Grav. 49 (2017) 69.

\bibitem{mr6} T. Harko, FSN Lobo, MK Mak, SV Sushkov, Phys. Rev. D 88 (2013) 044032.

\bibitem{mr7} T. Harko, FSN Lobo, MK Mak, SV Sushkov, Phys. Rev. D 88 (2013) 044032.

\bibitem{mr8} A. I. Qauli, M. Iqbal, A. Sulaksono, H.S. Ramadhan, Phys. Rev. D 93 (2016) 104056.

\bibitem{mr9} A. I. Qauli, A. Sulaksono, H. S. Ramadhan, I. Husin, arXiv:1710.03988.

\bibitem{mr10} B. Danila, T. Harko, F. S. N. Lobo, M. K. Mak, Phys. Rev. D 95 (2017) 044031.

\bibitem{ba1} E. Barausse, T. P. Sotiriou, J. C. Miller, Class. Quant. Grav. 25 (2008) 062001.

\bibitem{ba2} E. Barausse, T. P. Sotiriou, J C. Miller, Class. Quant. Grav. 25 (2008) 105008.

\bibitem{ba3} E. Barausse, T. P. Sotiriou, J C. Miller, EAS Publications Series 30 (2008) 189.

 \bibitem{pani} P.~Pani and T.~P.~Sotiriou,
  Phys.\ Rev.\ Lett.\  {\bf 109} (2012) 251102

\bibitem{ba4} Y-H. Sham, P. T. Leung, L-M. Lin, Phys. Rev. D 87 (2013) 061503.

\bibitem{ba5} J. Barrientos O. and G. F. Rubilar, Phys. Rev. D 93 (2016) 024021.

  \bibitem{gonzalo2} G.J. Olmo, Phys.Rev.D 78 (2008) 104026.
  
    \bibitem{kim} H-Ch. Kim, Phys. Rev. D 89 (2014) 064001.
  
  \bibitem{b6} A. Mana, L. Fatibene, M. Ferraris, JCAP 2015 (2015) 040.
  
  \bibitem{gd} G.J. Olmo, D. Rubiera-Garcia, arXiv:2007.04065.

  \bibitem{stab1} T-H. Sham, L-M. Lin, P. T. Leung, Phys. Rev. D 86 (2012) 064015.
  
  \bibitem{stab2} H. Sotani, Phys. Rev. D 89 (2014) 124037.
  
  \bibitem{stab3} M. Z. Bhatti, Z. Yousaf, Zarnoor, Gen. Rel. Grav. 51 (2019) 144.
  
  \bibitem{stab4} P. Pani, T. Delsate, and V. Cardoso, Phys. Rev. D 85 (2012) 084020.
  
  \bibitem{stab5} W-X. Feng, C-Q. Geng, L-W Luo, Chin. Phys. C 43 (2019) 083107.
 
  \bibitem{aneta2} A. Wojnar, Eur. Phys. J. C79 (2019) no.1, 51.
  
 \bibitem{gonzalo} G.J. Olmo, D. Rubiera-Garcia, A. Wojnar, Phys.Rev.D 100 (2019) 4, 044020.
 
    \bibitem{artur} A. Sergyeyev, A. Wojnar, Eur.Phys.J.C 80 (2020) 4, 313.  
    
     \bibitem{nn1} S. Banerjee, S. Shankar, T.P. Singh, JCAP 2017 (2017) 004.
 
 \bibitem{nn2} C. Wibisono, A. Sulaksono, Int. J. Mod. Phys. D, 27 (2018) 1850051.
 
 \bibitem{aneta3} A. Wojnar, Acta Phys.Polon.Supp. 13 (2020) 249.
       
     \bibitem{hansen} C. J. Hansen, S. D. Kawaler, V. Trimble, {\it Stellar Interiors. Physical Principles, Structure, and Evolution}, 
     2nd edition, Springer 2004.
     
     \bibitem{glen} N. K. Glendenning, {\it Compact Stars}, Springer 1996
         
    \bibitem{ewol} R. Kippenhahn, A. Weigert, A, Weiss, {\it Stellar structure and evolution}, Springer, 2012.
             
 \bibitem{hayashi} C. Hayashi, Publication of the Astronomical Society of Japan, Vol.13, 450-452 (1961).
         
     \bibitem{hll} L. Henyey, R. Lelevier, and R.D. Levee, Publications of the Astronomical Society of the Pacific 67.396 (1955): 154-160.
     
     \bibitem{hbg} L. Henyey, J.E. Forbes, N.L. Gould, The Astrophysical Journal 139 (1964): 306.
     
     \bibitem{hvb} L. Henyey, M.S. Vardya, P. Bodenheimer, The Astrophysical Journal 142 (1965): 841.
     
     \bibitem{szyd} M. Szydlowski, Marek, A.J. Maciejewski, Journal of Physics A: Mathematical and General 37.10 (2004): 3501.
      
  \bibitem{sak1} J. Sakstein, Phys. Rev. Lett. 115 (2015) 201101.
  
  \bibitem{sak2}   J. Sakstein, Phys. Rev. D 92 (2015) 124045.
  
  \bibitem{cris} M. Crisostomi, M. Lewandowski and F. Vernizzi, Phys. Rev. D 100 no.2, 024025 (2019).
        
    \bibitem{cant} M. Cantiello, J. Braithwaite, The Astrophysical Journal, 883: 106, 2019.
      
  \bibitem{haynak} C. Hayashi, T. Nakano, Progress of Theoretical Physics, 30(4), 460-474, (1963).
  
      \bibitem{morin} J. Morin et al., Monthly Notices of the Royal Astronomical Society 384.1 (2008): 77-86.
      
      \bibitem{Bildsten} L. Bildsten, E.F. Brown, C.D. Matzner, G. Ushomirsky, The Astrophysical Journal 482.1 (1997): 442.
  
  \bibitem{ush} G. Ushomirsky, C.D. Matzner, E.F. Brown, L. Bildsten, V.G. Hilliard, P.C Schroeder, The Astrophysical Journal 497 : 253-266, 1998.
     
  \bibitem{wright} N.J. Wright, et al, Monthly Notices of the Royal Astronomical Society 479.2 (2018): 2351-2360.
  
  \bibitem{aneta4} A. Wojnar, arXiv:2009.10983
      
  \bibitem{hill} L.A. Hillenbrand, R.J. White, The Astrophysical Journal 604.2 (2004): 741.
    
      \bibitem{tout} C.A. Tout, M. Livio, I.A. Bonnell, Mon. Not. R. Astron. Soc. 310, 360-376 (1999).
      
  \bibitem{banados} M. Banados and P. G. Ferreira,  Phys.Rev.Lett. 105, 011101 (2010).
  
   \bibitem{delsate2} T. Delsate and J. Steinhoff, Phys. Rev. Lett. 109, 021101 (2012).
    
\bibitem{BeltranJimenez:2017doy}
  J.~Beltran Jimenez, L.~Heisenberg, G.~J.~Olmo and D.~Rubiera-Garcia,
  Phys.\ Rept.\  {\bf 727} (2018) 1
     \bibitem{vol} . N. Vollick, Phys. Rev. D 69, 064030 (2004).
  
\bibitem{adria} A. Delhom-Latorre, G.J. Olmo, M. Ronco, Physics Letters B 780 (2018): 294-299.
 
  \bibitem{delf} X. Delfosse, T. Forveille, D. Ségransan, J.L. Beuzit, S. Udry, C. Perrier, M. Mayor, Astronomy and Astrophysics, 364, 217-224 (2000).
  
  \bibitem{mann1} A.W. Mann et al., The Astrophysical Journal, 804: 64, 2015.
  
  \bibitem{pars} S.G. Parsons et al.,  Monthly Notices of the Royal Astronomical Society, Vol. 481, Issue 1, 2018.
   
  \bibitem{mann2} A.W. Mann et al., The Astrophysical Journal, 871: 63, 2019.
  
    \bibitem{chab2} G. Chabrier et al., {\it The mass-radius relationship from solar-type stars to terrestrial planets: a review}, AIP Conference Proceedings. Vol. 1094. No. 1. American Institute of Physics, 2009.
  
  \bibitem{ser} A. Serenelli et al., arXiv:2006.10868.
  
  \bibitem{siess} L. Siess, E. Dufour, M. Forestini, Astron. Astrophys. 358: 593-599, 2000.

\end{thebibliography}
\end{document}